\documentclass[%
 reprint,
 preprintnumbers,
 amsmath, amssymb,
 aps,
 prl,
 showpacs,
]{revtex4-1}

\usepackage{graphicx}
\usepackage{dcolumn}
\usepackage{bm}
\usepackage{multirow}
\usepackage{threeparttable}
\usepackage{color}
\usepackage{ulem}

\begin{document}

\preprint{APS/123-QED}

\title{Quasiparticle Dynamics and Phonon Softening in FeSe Superconductors}

\author{C. W. Luo$ ^{1,*} $}
\author{I. H. Wu$ ^{1} $}
\author{P. C. Cheng$ ^{1} $}
\author{J.-Y. Lin$ ^{2} $}
\author{K. H. Wu$ ^{1} $}
\author{T. M. Uen$ ^{1} $}
\author{J. Y. Juang$ ^{1} $}
\author{T. Kobayashi$ ^{1,3} $}
\author{D. A. Chareev$ ^{4} $}
\author{ O. S. Volkova$ ^{5} $}
\author{ A. N. Vasiliev$ ^{5} $}

\affiliation{$ ^1 $Department of Electrophysics, National Chiao Tung University, Hsinchu 300, Taiwan}
\affiliation{$ ^2 $Institute of Physics, National Chiao Tung University, Hsinchu 300, Taiwan}
\affiliation{$ ^3 $Department of Applied Physics and Chemistry and Institute for Laser Science, University of Electro-Communications, 1-5-1Chofugaoka, Chofu, Tokyo 182-8585, Japan}
\affiliation{$ ^4 $Institute of Experimental Mineralogy, Chernogolovka, Moscow Region, 142432, Russia}
\affiliation{$ ^5 $Low Temperature Physics and Superconductivity Department, Moscow State University, 119991 Moscow, Russia}

\date{\today}

\begin{abstract}
Quasiparticle dynamics of FeSe single crystals revealed by dual-color transient reflectivity measurements (${\textrm{$\Delta$}} \textit{R}/\textit{R}$) provides unprecedented information on Fe-based superconductors. The amplitude of the fast component in ${\textrm{$\Delta$}} \textit{R}/\textit{R}$ clearly gives a competing scenario between spin fluctuations and superconductivity. Together with the transport measurements, the relaxation time analysis further exhibits anomalous changes at 90 and 230 K. The former manifests a structure phase transition as well as the associated phonon softening. The latter suggests a previously overlooked phase transition or crossover in FeSe. The electron-phonon coupling constant \textit{$\lambda$} is found to be 0.16, identical to the value of theoretical calculations. Such a small \textit{$\lambda$} demonstrates an unconventional origin of superconductivity in FeSe.
\end{abstract}

\maketitle

Since the discovery of Fe-based superconductors (FeSCs) in 2008 \cite{Y. Kamihara et al 2008}, tremendous experimental and theoretical effort has been devoted to exploring their characteristics. These Fe-based pnictide compounds exhibit a very interesting phase diagram, with antiferromagnetism (or spin-density wave) at low doping and superconductivity at intermediate doping \cite{Johnpierre Paglione et al 2010}. The simultaneous presence of magnetism and superconductivity in the phase diagram implies that magnetism plays an important role in the superconductivity mechanism. The existence of precursor superconductivity above \textit{T}$_{\textrm{c}}$ which competes with the spin-density wave order \cite{E. E. M. Chia et al 2010}, and a pseudogaplike feature with onset around 200 K \cite{T. Mertelj et al 2009} were observed on underdoped (Ba, K)Fe$_{\textrm{2}}$As$_{\textrm{2}}$ and nearly optimally doped SmFeAsO$_{\textrm{0.8}}$F$_{\textrm{0.2}}$, respectively. Additionally, a coherent lattice oscillation was also found in Co-doped BaFe$_{\textrm{2}}$As$_{\textrm{2}}$ using time-resolved pump-probe reflectivity with 40 fs time resolution \cite{B. Mansart et al 2009}. Among various FeSCs, the iron chalcogenide FeSe \cite{F. C. Hsu et al 2008} stands out due to its structure simplicity, which consists of iron-chalcogenide layers stacking one by another with the same Fe$^{\textrm{+2}}$ charge state as the iron pnictides. This so-called ``11'' system is so simple that it could be the key structure to understanding the origin of high-\textit{T}$_{\textrm{c}}$ superconductivity \cite{D. C. Johnston 2010}. There has been considerable concern over the interplay between electronic structure, phonons, magnetism, and superconductivity in 11-type FeSe. Therefore, further studies of their quasiparticle dynamics are indispensable to understanding the high-\textit{T}$_{\textrm{c}}$ mechanism in FeSCs. Here we report the first time-resolved femtosecond spectroscopy study of FeSe single crystals to elucidate the electronic structure and the quasiparticle (QP) dynamics. 

In this study, FeSe single crystals were grown in evacuated quartz ampoules using a KCl/AlCl$_{\textrm{3}}$ flux \cite{J.-Y. Lin et al 2010}. The crystalline structure of the samples was examined by x-ray diffraction. The superconducting transition temperature \textit{T}$_{\textrm{c}}$ of the FeSe single crystal was determined to be 8.8 K by the middle point of the resistive transition. The femtosecond spectroscopy measurement was performed using a dual-color pump-probe system (for light source, the repetition rate: 5.2 MHz, the wavelength: 800 nm, and the pulse duration: 100 fs) and an avalanche photodetector with the standard lock-in technique. The polarizations of the pump and probe beams were perpendicular to each other and parallel to the \textit{ab} plane of FeSe single crystals with no distinction between \textit{a} and \textit{b} orientation. The fluences of the pump beam and the probe beam are 9.92 and 1.40 $\mu$J/cm$^{2}$, respectively. The pump pulses have corresponding photon energy (3.1 eV) where the higher absorption occurred in the absorption spectrum of FeSe \cite{X. J. Wu et al 2007} and hence can generate electronic excitations. The QP dynamics is studied by measuring the photoinduced transient reflectivity changes (${\textrm{$\Delta$}} \textit{R}/\textit{R}$) of the probe beam with photon energy of 1.55 eV.

\begin{figure}
\begin{center}
\includegraphics[width=6.6cm]{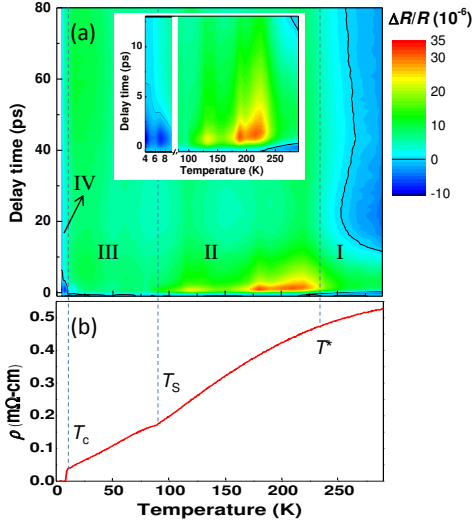}
\caption{\label{fig:fig1}(color online) (a) Temperature and delay time dependence of ${\textrm{$\Delta$}} \textit{R}/\textit{R}$ in an FeSe single crystal. The inset shows a part of the ${\textrm{$\Delta$}} \textit{R}/\textit{R}$ on an enlarged scale. Solid lines indicate ${\textrm{$\Delta$}} \textit{R}/\textit{R}$= 0. (b) Temperature dependence of the resistivity $  \rho$ indicates the high quality of an FeSe single crystal. The kink of $  \rho$($T$) at 90 K manifests the structure phase transition.}
\end{center}
\end{figure}

Electronic excitations generated by the pump pulses result in a swift rise of ${\textrm{$\Delta$}} \textit{R}/\textit{R}$ at zero time delay. The observed excitation is triggered by transferring the electrons from \textit{d} valence band of Fe to \textit{d} conduction band of Fe \cite{A. Subedi et al 2008}. At zero time delay, the number of the excited electrons generated by this nonthermal process is related to the amplitude of ${\textrm{$\Delta$}} \textit{R}/\textit{R}$. These high-energy electrons accumulated in the \textit{d} conduction band of Fe release their energy through the emission of longitudinal-optical (LO) phonons within several picoseconds \cite{F. S. Krasniqi et al 2008,Due to the 12}. The LO phonons further decay into longitudinal-acoustic (LA) phonons via anharmonic interactions, i.e., transferring energy to the lattice. This relaxation process can be detected using a probe beam as shown in Fig. 1(a), which shows the 2D ${\textrm{$\Delta$}} \textit{R}/\textit{R}$ taken on an FeSe single crystal. Four temperature regions appear. Above 230 K, \textit{T}* (region I), there is a fast negative response with a relaxation time of about 1.5 ps together with a long period of oscillation. When the temperature decreases into region II, a positive and slow response appears and ${\textrm{$\Delta$}} \textit{R}/\textit{R}$ gradually becomes smaller until \textit{T} = 90 K (\textit{T}$_{\textrm{s}}$) where the resistivity \textit{$\rho$}(\textit{T}) curve shows the kink [Fig. 1(b)]. Below 90 K (region III), the slow positive response disappears and is replaced by a complicated mixture of the positive and negative components as discussed later. For \textit{T} $<$ \textit{T}$_{\textrm{c}}$ (region IV), a long-lived negative response appears similar to the one in region I. These four regions in 2D ${\textrm{$\Delta$}} \textit{R}/\textit{R}$ are apparently corresponding to the feature of \textit{$\rho$}(\textit{T}) in FeSe. The relaxation processes ($\textit{t} >$ 0) of ${\textrm{$\Delta$}} \textit{R}/\textit{R}$ in FeSe can be phenomenologically described by

\begin{figure}
\begin{center}
\includegraphics*[width=8.8cm]{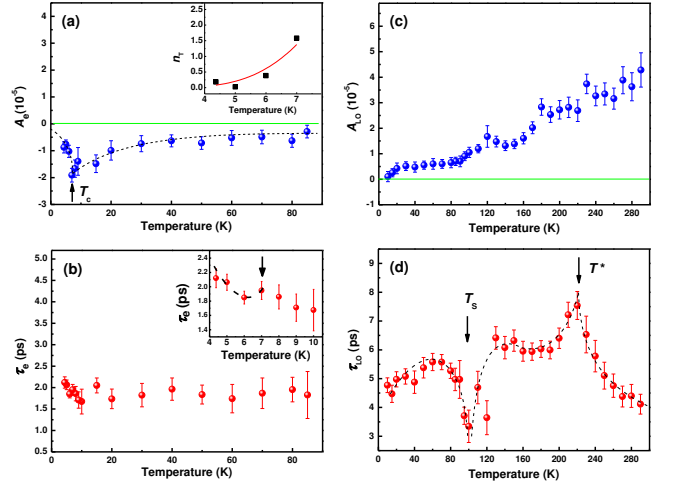}
\caption{\label{fig:fig2}(color online) Temperature dependence of the amplitude (a) \textit{A}$_{\textrm{e}}$, (c) \textit{A}$_{\textrm{LO}}$ and the relaxation time (b)$\tau_{\textrm{e}}$, (d) $\tau_{\textrm{LO}}$ by fitting Eq. (\ref{eq:one}). Inset of (a): Density of thermally excited QPs, $n_{\textrm{T}}(T)$. A solid line fits to the Rothwarf-Taylor model. Inset of (b): The temperature-dependent $\tau_{\textrm{e}}$ is shown between 4 and 10 K. Dashed lines are guides to the eyes. The arrow indicates the divergence of $\tau_{\textrm{e}}$ around \textit{T}$_{\textrm{c}}$.}
\end{center}
\end{figure}

\begin{eqnarray}
\label{eq:one}
\frac{{\textrm{$\Delta$}} \textit{R}}{\textit{R}} = \textit{A}_{\textrm{e}}e^{-\textit{t}/\tau_{\textrm{e}}} + \textit{A}_{\textrm{LO}}e^{-\textit{t}/\tau_{\textrm{LO}}} + \textit{A}_{\textrm{0}} \nonumber\\ + \textit{A}_{\textrm{LA}}e^{-\textit{t}/\tau _{\textrm{LA}}}\sin[2\pi t/T(t) + \phi]
\end{eqnarray}

The first term in the right-hand side of Eq. (\ref{eq:one}) is the decay of the excited electrons with an initial population number, \textit{A}$_{\textrm{e}}$, and a relaxation time, $\tau_{\textrm{e}}$. The second term is the phonon number, \textit{A}$_{\textrm{LO}}$, and a corresponding decay time, $\tau_{\textrm{LO}}$. The third term describes energy loss from the hot spot to the ambient environment within the time scale of microsecond, which is far longer than the period of the measurement ($ \sim $ 150 ps) and hence is taken as a constant. The last term is the chirped oscillation component associated with strain pulse propagation \cite{This oscillation is 13}: \textit{A}$_{\textrm{LA}}$ is the amplitude of the oscillation; $\tau_{\textrm{LA}}$ is the damping time; \textit{T}(\textit{t}) is the time-dependent period; \textit{$\phi$} is the initial phase of the oscillation.

We first discuss the temperature-dependent ${\textrm{$\Delta$}} \textit{R}/\textit{R}$ in FeSe. Fig. \ref{fig:fig1}(a) shows ${\textrm{$\Delta$}} \textit{R}/\textit{R}$ from 290 down to 4.4 K, from which each component described above can be extracted using Eq. (\ref{eq:one}). Results of the extraction are shown in Fig. \ref{fig:fig2}(a)-2(d). For the fast component in ${\textrm{$\Delta$}} \textit{R}/\textit{R}$, a remarkable increase in the amplitude (\textit{A}$_{\textrm{e}}$) is shown in Fig. \ref{fig:fig2}(a) when \textit{T} decreases from 90 K. The temperature dependence of spin fluctuations below \textit{T}$_{\textrm{s}}$ in FeSe which were revealed by $^{77}$Se NMR measurements \cite{T. Imai et al 2009} is totally different from that in cuprates \cite{K. S. Bedell 16}. Therefore, this temperature-dependent \textit{A}$_{\textrm{e}}$ below \textit{T}$_{\textrm{s}}$ is attributed to the strong antiferromagnetic spin fluctuations which may associate with the structural transformation at 90 K rather than with the pseudogap in cuprates. Intriguingly as \textit{T} $<$ \textit{T}$_{\textrm{c}}$ (region IV in Fig. \ref{fig:fig1}), \textit{A}$_{\textrm{e}}$ decreases due to the onset of superconductivity. This suggests that the spin fluctuations and superconductivity are competing factors in the FeSe system. The above results provide the first compelling experimental evidence of competing orders in FeSe, which are consistent with the theoretical calculations \cite{Hongliang Shi et al 2011}. It is noted that the experimental evidence on the competing orders was also reported in underdoped (Ba, K)Fe$_{\textrm{2}}$As$_{\textrm{2}}$ \cite{E. E. M. Chia et al 2010}. According to the Rothwarf-Taylor model \cite{A. Rothwarf et al 1967}, the density of thermally excited QPs \textit{n}$_{T}\propto$ [\textit{A}$_{\textrm{e}}$(\textit{T})/\textit{A}$_{\textrm{e}}$(\textit{T}$\rightarrow$0)] - 1 as shown in the inset of Fig. \ref{fig:fig2}(a). The temperature-dependent behavior of \textit{n}$_{T}$ can be further fitted by \textit{n}$_{T}\propto$ [$\triangle$(\textit{T})\textit{T}]$ ^{1/2} $exp[-$\triangle$(\textit{T})/\textit{T}], where $\triangle$(\textit{T}) is the superconducting energy gap. Assuming a BCS temperature dependent $\triangle$(\textit{T}) = $\triangle$(0)[1-(\textit{T}/\textit{T}$_{\textrm{c}}$)]$ ^{1/2} $, the fits lead to $\triangle$(0) = 3.24\textit{k}$_{B}$\textit{T}$_{\textrm{c}}$ in accord with the value obtained from specific heat measurements \cite{J.-Y. Lin et al 2010}. [This isotropic $\triangle$(\textit{T}) scenario is certainly too simplified, but it suffices to estimate the scale of $\triangle$(\textit{T}).]

The presence of a gap in the QP density of states gives rise to a bottleneck for carrier relaxation, which is clearly observed in the relaxation time $\tau_{\textrm{e}}$ close to \textit{T}$_{\textrm{c}}$. The mechanism of the bottleneck can be described by the Rothwarf-Taylor model \cite{A. Rothwarf et al 1967}. When two QPs with energies higher than $\triangle$, a high energy boson (HEB) with energy $ \omega\geq2\triangle $ is created. The HEBs that remain in the excitation volume subsequently break additional Cooper pairs, effectively preventing QP from recombination. Till $ \omega<2\triangle $ and the Cooper pairs could not be broken further by HEBs, the number of QPs finally decreases in several picoseconds. In the case of a mean-field-like gap, i.e., the gap gradually shrinks with \textit{T} approaching \textit{T}$_{\textrm{c}}$ from the low-temperature side, more HEBs are available to regenerate QPs. The recombination processes become less and less efficient. Hence, $\tau_{\textrm{e}}$ below \textit{T}$_{\textrm{c}}$ through Cooper pair recombination is longer than 1.5-2 ps and diverges around \textit{T}$_{\textrm{c}}$, which is dominated by the superconducting gap opening. To compare, $\tau_{\textrm{e}}$ is $\sim$ 1.5-2 ps between \textit{T}$_{\textrm{c}}$ and 90 K and is dominated by the appearance of spin fluctuations in this region.

For the slower component, the amplitude \textit{A}$_{\textrm{LO}}$ monochromatically decreases as \textit{T} does, and then it completely disappears in the superconducting state. On the contrary, the relaxation time $\tau_{\textrm{LO}}$ exhibits two marked anomalies both at \textit{T}*$ \sim $ 230 K and \textit{T}$_\textrm{s}\sim $ 90 K, corresponding to the boundary between region I and II and between region II and III, respectively. During the structural transition from a tetragonal phase to an orthorhombic phase \cite{T. M. McQueen et al 20091}, the energy of LO phonons more efficiently releases to lattice and results in a significantly shorter relaxation time $\tau_{\textrm{LO}}$ at \textit{T}$_{\textrm{s}}$. On the other hand, the abnormally long relaxation time $\tau_{\textrm{LO}}$ around 230 K suggests an elusive higher-temperature phase transition or crossover to undermine the energy release efficiency. The sign change of Seebeck coefficient of FeSe was found to be also at \textit{T}* \cite{T. M. McQueen et al 20092}.

By fitting the ${\textrm{$\Delta$}} \textit{R}/\textit{R}$ curves with Eq. (\ref{eq:one}), dynamic information on QPs and phonons is available, which includes the number of QPs, the relaxation time of QPs, and the energy of phonons. In a metal, the photoinduced QP relaxation time is governed by transferring energy from electron to phonon with electron-phonon coupling strength $\lambda$ \cite{P. B. Allen 1987}.

\begin{equation}
\label{eq:two}
\dfrac{1}{\tau_{\textrm {e}}}=\dfrac{3\hbar\lambda\langle\omega^{2}\rangle}{\textrm{$\pi$}{k_{\textrm {B}}}T_{\textrm {e}}}
\end{equation}

\begin{table*}
\begin{center}
\tabcolsep=0.22cm
\caption{The parameters for $\lambda$ estimated at \textit{T}= 20 K of an FeSe single crystal.}
\begin{threeparttable}
\begin{tabular}[c]{cccccccccc}
\hline
 & \textit{T}$_{\textrm{c}}$ (K) & \textit{R} (400 nm) & \textit{F} ($\mu {\textrm{J/cm}^{2}}$)  & $\gamma^{\textrm{a}}$ (mJmol$^{-1}$K$^{-2}$) &  $\tau_{\textrm{e}}$ (ps)  & $A_{\textrm{1g}}^{\textrm{b}}$ (meV) &  $\lambda$$ \left\langle \omega^{2}\right\rangle $$^{\textrm{c}}$ (meV$^{2}$)  & $\lambda$ \tabularnewline
\hline
FeSe & 8.8 & 0.25 & 9.92 & 5.73 & 1.75 & 19.9 & 61.3 & 0.16\tabularnewline
\hline
\end{tabular}
\begin{flushleft}
\begin{tablenotes}
\item $^{\textrm{a}}$From Ref. \cite{J.-Y. Lin et al 2010}. $^{\textrm{b}}$From Ref. \cite{P. Kumar et al 2010}. $^{\textrm{c}}$Obtained from Eq. (\ref{eq:two}).
\end{tablenotes}
\end{flushleft}
\end{threeparttable}
\end{center}
\end{table*}

\begin{figure}
\begin{center}
\includegraphics*[width=8cm]{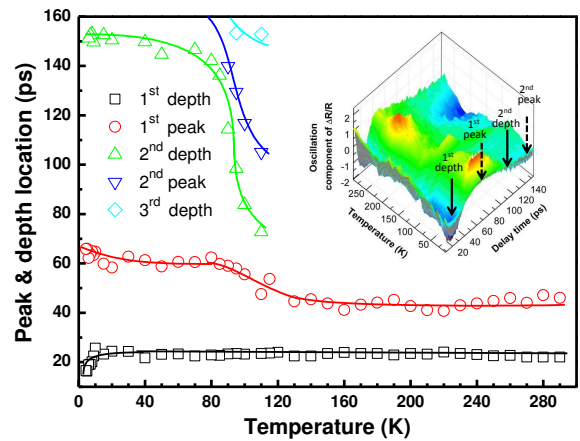}
\caption{\label{fig:fig3}(color online) Temperature-dependent peaks and depths of the oscillation component in ${\textrm{$\Delta$}} \textit{R}/\textit{R}$. Inset:  Temperature and delay time dependence of the 3D oscillation component was obtained by subtracted the decay background [the 1$^{\textrm{st}}$, 2$^{\textrm{nd}}$, and 3$^{\textrm{rd}}$ terms in Eq. (\ref{eq:one})] from ${\textrm{$\Delta$}} \textit{R}/\textit{R}$ of Fig. \ref{fig:fig1}(a). Solid lines are guides to the eyes.}
\end{center}
\end{figure}

\noindent where $\lambda\langle\omega^{2}\rangle$ is the second moment of the Eliashberg function and \textit{T}$_{\textrm{e}}$ can be further described by \cite{D. Boschetto et al 2008}

\begin{equation}
\label{eq:three}
T_{\textrm {e}}=\langle\sqrt{T_{\textrm {i}}^{2}+{\dfrac{2(1-R)F}{l_\textrm {s}\gamma}}e^{-z/l_\textrm {s}}}\rangle
\end{equation}

\noindent where \textit{R} is the unperturbed reflectivity at 400 nm, \textit{F} is the pumping fluences and $\gamma$ is the linear coefficient of heat capacity due to the electronic subsystem. The mean value is taken for the depth \textit{z} going from the crystal surface down to the skin depth \textit{l}$_{\textrm{s}}\sim$ 24 nm (estimated from the skin depth of electromagnetic wave in metal, $ \lambda/4\textrm {$\pi$}k $). All of the parameters for the calculations of electron-phonon coupling strength were listed in Table I. For the estimate of $\langle\omega^{2}\rangle$, some vibrational modes are more efficiently coupled to QPs than others are. In the case of Co-doped BaFe$_{\textrm{2}}$As$_{\textrm{2}}$, the symmetric A$_{\textrm{1g}}$ mode is coherently excited by photoexcitation and efficiently coupled \cite{B. Mansart et al 2009}. Consequently, we take the A$_{\textrm{1g}}$ mode into account in the present case of FeSe, which is the strongest phonon mode in electron-phonon spectral function, $ \alpha^{\textrm{2}}F(\omega) $ \cite{A. Subedi et al 2008}. By Eq. (\ref{eq:two}), the consequent electron-phonon (A$_{\textrm{1g}}$ mode) coupling constant is $\lambda$ = 0.16 in FeSe. This value is consistent with the theoretical results of $\lambda$ = 0.17 \cite{A. Subedi et al 2008} obtained by using linear response within the generalized gradient approximation (GGA). Furthermore, we can use the McMillan formula \textit{T}$_{\textrm{c}}$=($ \langle\omega\rangle $/1.2)exp$\lbrace{-[1.40(1+\lambda)]/[\lambda-\mu^{*}(1+0.62\lambda)]}\rbrace$ to evaluate the critical temperature \textit{T}$_{\textrm{c}}$ \cite{W. L. McMillan 1968}. Taking $ \langle\hbar\omega\rangle $= 19.9 meV and $ \mu $*= 0, we obtain \textit{T}$_{\textrm{c}}\sim$ 0.08 K, which is one order of magnitude lower than the actual \textit{T}$_{\textrm{c}}$ of about 8.8 K. Actually, the \textit{total} coupling strength between quasiparticles and bosons is estimated to be 1.55 from the specific heat measurements \cite{J.-Y. Lin et al 2010}. Therefore, the present study experimentally verifies that the superconductivity in FeSe should be attributed to the unconventional mechanism other than electron-phonon coupling.

Further insight into the phase transition in FeSe is provided by the study of the oscillation component of ${\textrm{$\Delta$}} \textit{R}/\textit{R}$. Temperature dependence of a strain pulse (LA phonons) propagation was clearly observed in the oscillation feature of ${\textrm{$\Delta$}} \textit{R}/\textit{R}$ after subtracting the decay background [i.e., the 1$^{\textrm{st}}$, 2$^{\textrm{nd}}$, and 3$^{\textrm{rd}}$ terms in  Eq. (\ref{eq:one})], as shown in the inset of Fig. \ref{fig:fig3}. In order to quantify the properties of LA phonons in FeSe, the locations of the peaks and depths of the oscillation in the inset of Fig. \ref{fig:fig3} are shown as the function of temperatures in Fig. \ref{fig:fig3}. At high temperatures, the damping time is very short and the oscillation sustains only for one period. However, the number of oscillation periods significantly increases below 100 K; hence the damping time becomes much longer. This means that the LA phonons can propagate further into the interior of FeSe crystals with the orthorhombic structure. According to the difference between 1$^{\textrm{st}}$ depth (at 23.52 ps) and 2$^{\textrm{nd}}$ depth (at 72.78 ps) at \textit{T}= 110 K, the phonon frequency is found to be 20.3 GHz. The phonon energy is estimated to be $ \sim $ 0.087 meV. The coherent acoustic phonon detected by a pump-probe reflectivity measurement can be described as a Brillouin scattering \cite{L. Brillouin 1922} phenomenon occurring in the materials after the excitation of pump pulses. The scattering condition is $q_{\textrm {phonon}}=2nk_{\textrm {probe}}\cos(\theta_{\textrm {i}})$, where \textit{q}$_{\textrm{phonon}}$ is the phonon wave vector, \textit{n} is the real part of the refractive index, and the probe photon has a wave vector \textit{k}$_{\textrm{probe}}$ arriving at an incident angle \textit{$\theta$}$_{\textrm{i}}$ (inside crystals) with respect to the surface normal. Following this scattering condition, the probe beam acts as a filter to select the acoustic wave propagating along the scattering plane symmetry axis, i.e., the normal to the crystal surface and traveling with the wave vector \textit{q}$_{\textrm{phonon}}$. The energy of the acoustic wave is $\textit{E}_{\textrm {phonon}}= \hbar{\omega}_{\textrm {phonon}}= \hbar{q}_{\textrm {phonon}}v_{\textrm {s}}= \hbar{2n}v_{\textrm {s}}k_{\textrm {probe}}\cos (\theta_{\textrm {i}})$, where \textit{v}$_{\textrm{s}}$ is the sound velocity along normal direction of crystal surface. Using \textit{$\lambda$}$_{\textrm{probe}}$= 800 nm, \textit{n}$_{\textrm{probe}}$= 2 \cite{X. J. Wu et al 2007}, \textit{$\theta$}$_{\textrm{i}}$= 2.5 $ ^{o} $ [estimated from the incident angle (5$ ^{o} $) of the probe beam by Snell's law] and \textit{v}$_{\textrm{s}}$= 3.58 km/s \cite{S. Chandra 2010}, the phonon energy, \textit{E}$_{\textrm{phonon}}$, is calculated to be 0.077 meV, which is very close to the result, 0.087 meV, directly obtained from the above ${\textrm{$\Delta$}} \textit{R}/\textit{R}$ measurements.

\begin{figure}
\begin{center}
\includegraphics*[width=7cm]{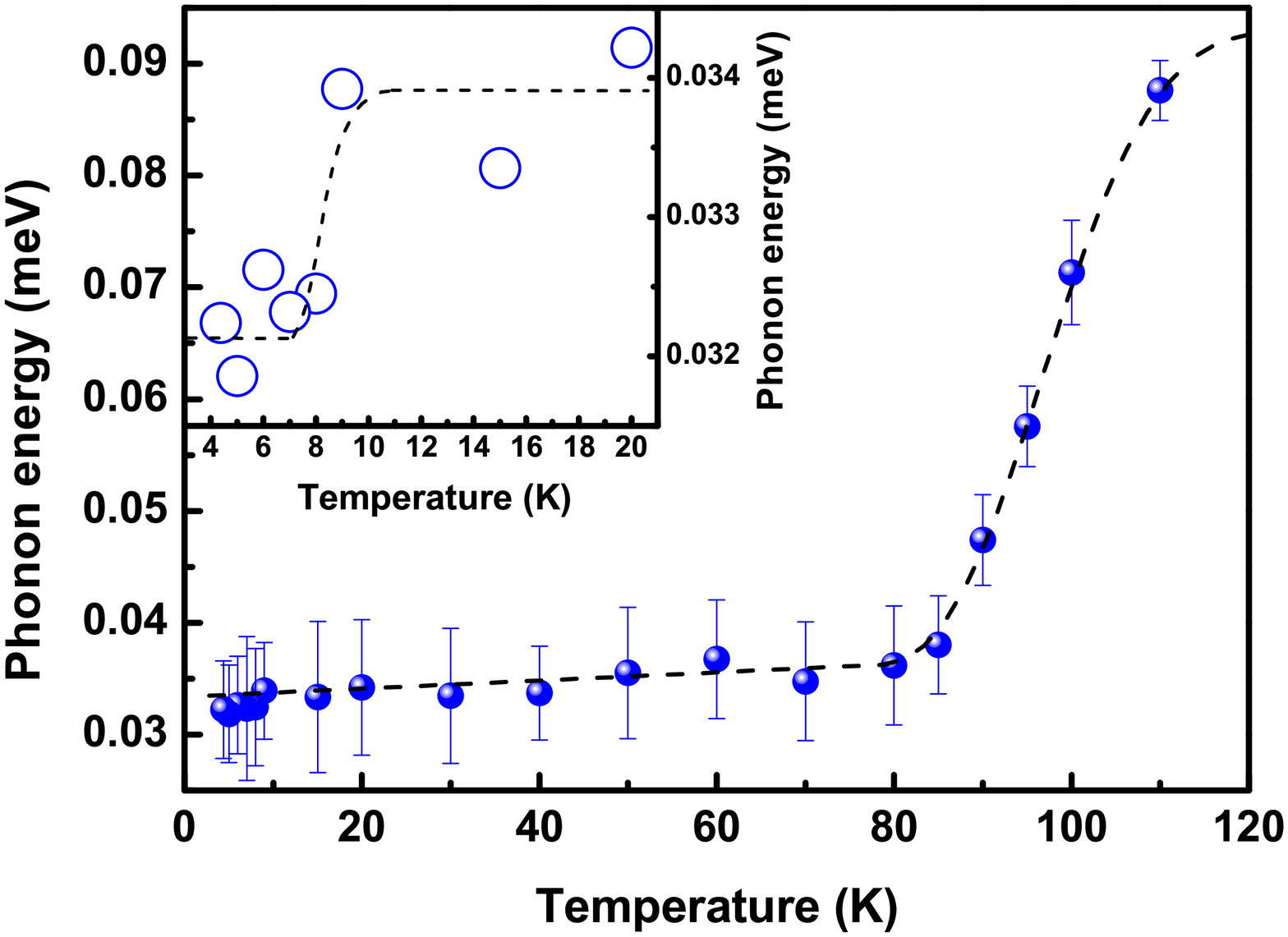}
\caption{\label{fig:fig4}(color online) Temperature dependence of the phonon energy derived from the oscillation component in Fig. \ref{fig:fig3}. The inset shows a part of the temperature-dependent phonon energy on an enlarged scale. Dashed lines are guides to the eyes.}
\end{center}
\end{figure}

We further investigate the temperature dependence of the LA phonon energy as shown in Fig. \ref{fig:fig4}. The phonon energy dramatically drops by 60$ \% $ \cite{By using k 28} around 90 K where a structural phase transition occurs, and then keeps constant at low temperatures. Additionally, we found that the LA phonons also soften by 6$ \% $ in the superconducting state as shown in the inset of Fig. \ref{fig:fig4}, which is consistent with the larger distance between 1$^{_{\textrm{st}}}$ depth and 1$^{_{\textrm{st}}}$ peak in Fig. \ref{fig:fig3}. Very recently, this softening of lattice has been found in Co-doped BaFe$_{\textrm{2}}$As$_{\textrm{2}}$ by resonant ultrasound spectroscopy \cite{R. M. Fernandes et al 2010, T. Goto et al 2011}. Fernandes \textit{et al}. \cite{R. M. Fernandes et al 2010} found the 16$ \% $ softening of shear modulus in BaFe$_{\textrm{1.84}}$Co$_{\textrm{0.16}}$As$_{\textrm{2}}$ at \textit{T}$_{\textrm{c}}$= 22 K. For the nonsuperconducting case of BaFe$_{\textrm{2}}$As$_{\textrm{2}}$, however, the rather large softening of 90$ \% $ was observed around 130 K, which is the structural and antiferromagnetism phase transition temperature. Similarly, a large phonon softening due to structural phase transition and a rather small phonon softening due to the superconductive phase transition was also observed in 11-type FeSe in this Letter. These results suggest that the reduction of phonon energy at both the structural and the superconducting phase transition is a general feature in FeSCs. The above phonon softening may participate in the superconducting pairing, albeit not the mechanism responsible for high \textit{T}$_{\textrm{c}}$ in FeSCs.

In summary, we have studied the ultrafast quasiparticle dynamics and phonon softening in FeSe single crystals by dual-color femtosecond spectroscopy. The temperature dependence of the amplitude \textit{A}$_{\textrm{e}}$ suggests a competing scenario between spin fluctuations and superconductivity. The relaxation time $\tau_{\textrm{e}}$ of ${\textrm{$\Delta$}} \textit{R}/\textit{R}$ reveals an electron-phonon coupling constant $ \lambda $= 0.16. The structure phase transition at 90 K is elucidated through both the electrical transport properties and the anomalous changes of relaxation time $\tau_{\textrm{LO}}$ of ${\textrm{$\Delta$}} \textit{R}/\textit{R}$. A previously unknown feature at 230 K is further identified. Moreover, the energy of LA phonons at 110 K was estimated to be 0.087 meV from the oscillation component of ${\textrm{$\Delta$}} \textit{R}/\textit{R}$, which markedly softens around both the structural phase transition and superconducting transition. Our results provide vital understanding of the competing picture between the spin fluctuations and superconductivity and of the role of phonons in Fe-based superconductors.

This project was financially sponsored by the National Science Council (Grants No. NSC 98-2112-M-009-008-MY3 and No. NSC 98-2112-M-009-005-MY3) and the Ministry of Education (MOE-ATU plan at National Chiao Tung University). This work was also supported by the Russian Ministry of Science and Education under Grant No. 11.519.11.6012. We thank Professor B. L. Young for discussions and Y. S. Hsieh for technical support.


\begin{thebibliography}{999}
\bibitem{This oscillation is 13}This oscillation is induced by electronic and lattice stresses. In isotropic medium, the thermoelastic stress can be expressed as $ \sigma_{\textrm{p}}= -3K\beta{\textrm{$\Delta$}}\textit{T}_{\textrm{l}}$, where \textit{K} is the bulk elastic modulus, ${\textrm{$\Delta$}}\textit{T}_{\textrm{l}}$ is the lattice temperature rise and $\beta$ is the linear thermal-expansion coefficient. In Thomen's model [14], the oscillation of ${\textrm{$\Delta$}} \textit{R}/\textit{R}$ is caused by the interference between the probe beams reflected from the crystal surface and the wave front of the propagating strain pulse with the modulated refractive index.
\end{thebibliography}
\end{document}